\documentclass[prl,twocolumn,10pt,showpacs]{revtex4}
\date{\today}

\newcommand{\insertplot}[5]{\begin{figure}
 \hfill\hbox to 0.05in{\vbox to #5in{\vfill
 \inputplot{#1}{#4}{#5}}\hfill}
 \hfill\vspace{-.1in}
 \caption{#2}\label{#3}
 \end{figure}}
 \newcommand{\inputplot}[3]{
 \special{ps: plotfile #1}
}
\newcounter{fig}   

\usepackage{epsfig}
\usepackage{amsmath}
\usepackage{amsfonts}
\usepackage{graphicx}
\usepackage{amssymb}
\usepackage{ifthen}

\begin{document}

\title{Could the primordial radiation be responsible for vanishing of topological defects?}
\author{Tomasz Roma\'nczukiewicz}
\affiliation{Institute of Physics, Jagiellonian University, Krak\'ow, Poland}

\begin{abstract}
We study the motion of topological defects in 1+1 and 2+1 d relativistic $\phi^6$ model with three equal vacua in the presence of radiation. We show 
that even small fluctuations can trigger a chain reaction leading to vanishing of topological defects. Only one vacuum remains stable and domains 
containing other vacua vanish. We explain this phenomenon in terms of radiation pressure (both positive and negative). We construct an effective 
model which translates the fluctuations into additional term in the field theory potential. In case of two dimensional model we find a relation 
between the critical size of the bulk and amplitude of the perturbation.
\end{abstract}

\maketitle

\noindent
{\it ~Introduction.~}
Topological defects arise in surprisingly many branches of physics. They can be found in liquid crystals \cite{Lavrentovich}, liquid helium 
\cite{Vollhardt}, ferromagnets, superconductors, graphene \cite{Yuri} and many more important physical substances. 
It is also natural to expect that topological defects should have been created in large numbers in the early Universe via Kibble-Zurek mechanism 
during some symmetry breaking phase transitions \cite{TWKibble1,TWKibble2}. 
Unfortunately there are no direct observation evidence proving such objects ever existed. However, it might be plausible that some linear defects 
(cosmic strings) could give origins to some large scale structures in the Universe. Some observed fluctuations in the cosmic microwave background  
referred to as ,,cold spot'' 
could be explained as remnants of textures from early stages the Universe \cite{coldSpot1,ColdSpot2}. Topological defects are sometimes also 
considered as one of the dark matter candidates \cite{darkPeter}.
Surprisingly, there are no signs of other defects like monopoles and domain walls. 

Topological defects can interact with each other as well as with some other objects like oscillons \cite{Hindmarsh, trom4}. A very interesting 
interaction also can be observed between topological defects and radiation. In some cases the radiation can exert an ordinary radiation pressure 
proportional to the square of amplitude of incident wave. However, some defects, like kinks in a very popular $\phi^4$ theory, are transparent to the 
radiation in the first order \cite{trom1, trom2}. Sine-Gordon kinks are exactly transparent in all orders. Higher order analysis in $\phi^4$ model 
revealed a surprising feature. The kinks undergo the negative radiation pressure (NRP) which accelerates the kinks towards the source of radiation. 
The acceleration of a kink in $\phi^4$ model is proportional to the fourth power of amplitude of the wave. In models with two scalar fields with 
different masses it is possible that the radiation can exert both positive and negative radiation pressure depending on the composition of the wave 
\cite{trom5, tromtractor}. In such a case the force exerted on the kink is proportional to the square of the amplitude.
More recently some other examples were discussed in case of light and matter waves which, when scattered on a small objects, can bend in such 
a way that the object would feel the pulling force \cite{Ruffner, Novitzky2}.
Mixing between different frequencies can cause the NRP in case of solitons with rotating phase as in Coupled Nonlinear Schr\"odinger Equation 
\cite{trom7}.
We want to emphasize that the NRP seen in case of solitons in the present and our previous papers is of a very different nature 
then the one described in optical physics. It can be exerted on flat and infinite surfaces where simple bending of the light or other wave 
trajectories is not an option.

In the present paper we consider a mechanism which could increase the rate of the domain wall collisions. 
In models with at least two equal minima of the potential but with different masses of small perturbations around those vacua. 
The interest in such models has increased recently \cite{trom6, ChristovIvan,Bazeia} but they were considered many years ago as for example so called 
bag models of hadrons \cite{bagmodel}.
Static kinks (or domain walls in general) in such models have asymmetric profiles. 
We show that despite the fact that the vacua have the same energies (they are 
true vacua) the kinks usually accelerate in one direction no matter from where the radiation comes. 
Antykinks accelerate in opposite direction. Any small perturbation can therefore trigger a chain reaction during which defects collide and create 
more radiation accelerating other defects causing more collisions.

The present letter is organized as follows. First we define our example $\phi^6$ model, than we show how kinks interact with monochromatic wave in 
case of two vacua with different masses of scalar field. 
We derive an analytic formula for the force with which such monochromatic wave acts on a kink. Next we show how generic perturbation can influence 
the stability of kink system (a lattice). 
In particular we study the effect of random  fluctuations with Gaussian distribution which could simulate the primordial radiation in the early 
Universe. We show that the fluctuations are in some ways equivalent to the shift of the vacua. We also compare the results with other models.
The last section concerns higher dimensional case. We find a critical size of a circular domain wall which could either grow or collapse depending on 
what type of vacuum is inside and how large the fluctuations are.

\begin{figure}
 \begin{center}
 \includegraphics[width=\linewidth]{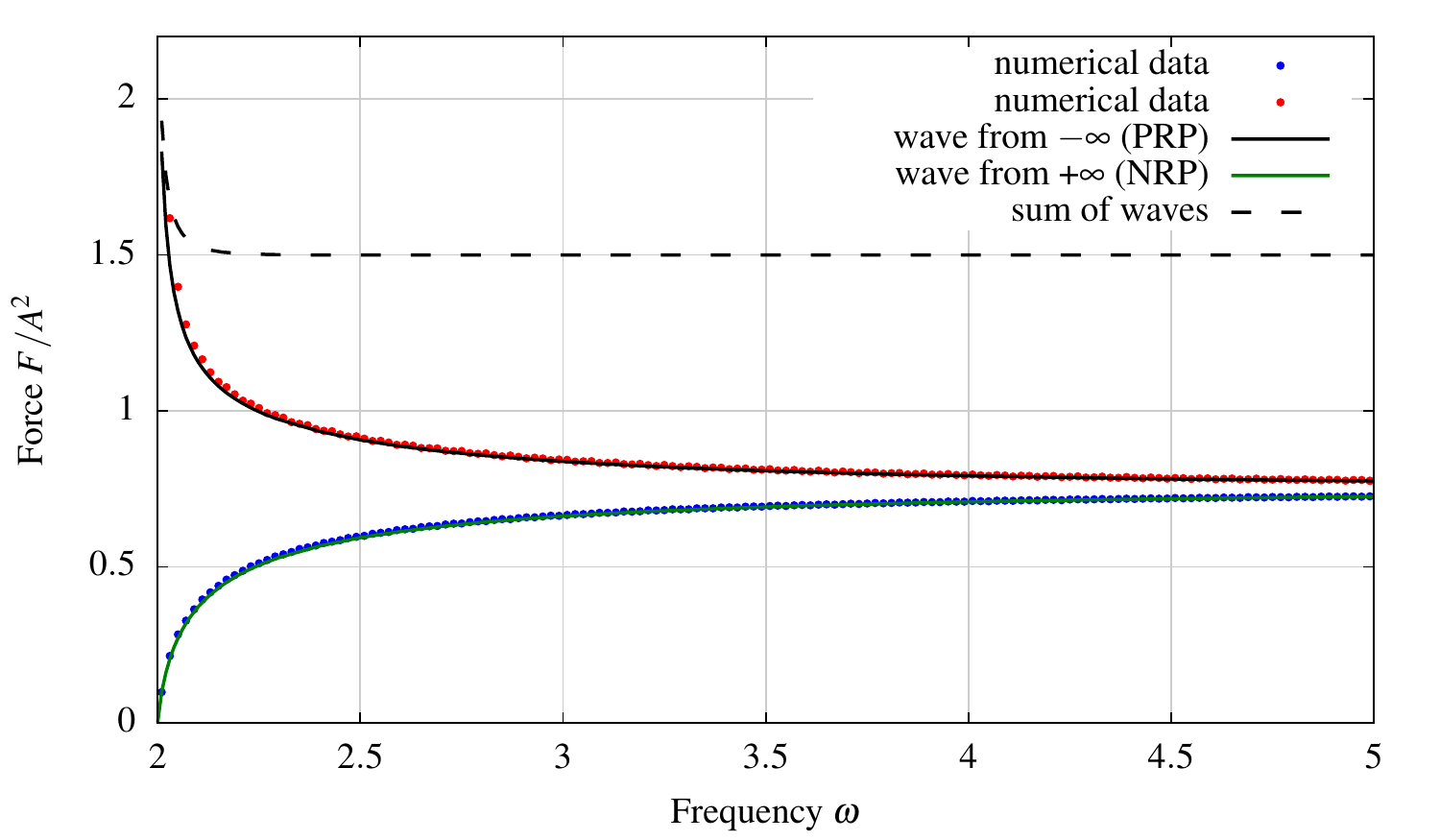}
 \caption{\small
 Theoretical values of the force exerted on the kink. In both cases the force is positive. The color points are the results of numerical calculations 
for $\mathcal{A}=0.05$.\\[-30pt]}\label{fig:1}
\end{center}
\end{figure}

{\it ~The model.~}
In the present paper we consider  one and two dimensional $\phi^6$ theory, which can be defined by the
rescaled Lagrangian density \cite{Lohe:1979mh,trom6}
\begin{equation}
   \label{Lag}
{\cal L} = \frac{1}{2}\partial_\mu \phi \partial^\mu \phi - 
\frac{1}{2}\phi^2\left(\phi^2 -1\right)^2.
\end{equation} 
The model has three vacua $\phi_v \in \{-1,0,1\}$. Small perturbations around these vacua have different masses: $m_0=1$ and $m_1=2$ for $\phi_v=0$ 
and $\phi_v=\pm1$ respectively.
In one dimensional case the kinks and antykinks can be found from a single solution
$\phi_K(x) \equiv \phi_{(0,1)}(x)= \sqrt{(1{+}\tanh x)/2}$
using the discrete symmetries of the model:
The masses of all kinks are $M=1/4$.
Small perturbation added to the kink solution $\phi(x,t) = \phi_s(x) + \eta(x) e^{i\omega t}$ is governed by a linearized equation 
$-\eta_{xx} + U(x) \eta = 
\omega^2 \eta$ where 
the potential $U(x)$ is
\begin{equation}
 \label{Pot}
U(x) = 15 \phi_s^4 - 12 \phi_s^2 +1.
\end{equation} 
Note that when $x\to-\infty$ the potential $U\to1$ and $U(x\to\infty)\to4$.
Solutions to this linearized equation can be found in analytic form in \cite{Lohe:1979mh}. 
Let us consider a wave traveling from the left side of kink \textit{i.e.}\ from $\phi=0$ vacuum.
Asymptotic form of these solutions for frequencies above the two mass thresholds due to Lohe can be written as
\begin{equation}
   \begin{cases}
      \eta_{+\infty}(x)\to e^{ikx}/A(q,k),\\
      \eta_{-\infty}(x)\to e^{iqx}+\frac{A(-q, k)}{A(q,k)}e^{-iqx}
   \end{cases}
\end{equation} 
with
\begin{equation}
   \begin{gathered}
   q=\sqrt{\omega^2-1},\;k=\sqrt{\omega^2-4},\\
   A(q,k) = \frac{\Gamma(1-ik)\Gamma(-iq)}{\Gamma(-\frac{1}{2}ik-\frac{1}{2}iq+\frac{5}{2})\Gamma(-\frac{1}{2}ik-\frac{1}{2}iq-\frac{3}{2})}.
   \end{gathered}
\end{equation} 
This solution represents a wave traveling from $-\infty$ with amplitude $1$. Amplitude of the reflected wave is equal to $\frac{A(-q, k)}{A(q,k)}$, 
and amplitude of the transient wave is $\frac{1}{A(q,k)}$. 
We can use this form to calculate the momentum and energy balance far away from the kink. 
From Noether's theorem, the conservation laws for energy and momentum density can be written as:
\begin{subequations}
\begin{align}
\partial_t\mathcal{E} &= \partial_x\left(\phi'\dot\phi\right),\\
\partial_t\mathcal{P} &= -\frac{1}{2}\partial_x\left(\dot\phi^2+\phi'^2-2U(\phi)\right).
\end{align}
\end{subequations}
Integrating the above expressions inside interval $[-L,L]$ and averaging over a period $T$ we obtain energy and momentum balance just using 
asymptotic form of scattering solutions.
If the kink does not move initially the conservation laws give (for $\phi=\phi_s+\mathcal{A}\mathop{\text{re}}(e^{-i\omega 
t}\eta(x))$):
\begin{equation}
   F_{+\infty}(q,k) = \frac{1}{2}\frac{\mathcal{A}^2}{|A(q,k)|^2}\left(2|A(-q,k)|^2q^2+qk-k^2\right).
\end{equation} 
 We can perform a very similar calculation for the second case when the wave is coming from $+\infty$. The force in this case can be expressed as:
\begin{equation}
   F_{-\infty}(q,k) = -F_{+\infty}(k,q)\equiv\mathcal{A}^2(\omega)f(\omega). \label{eq:forza}
\end{equation} 
Fig \ref{fig:1} shows the force in both cases. Note that the force is positive for all frequencies. The kink will always accelerate towards $+\infty$ 
no matter which direction the wave comes from. In the first case the wave comes from $\phi=0$ ($m=1$) and exerts positive radiation pressure. In the 
second case it comes from the second vacuum $\phi=1$ with mass $m=2$ and exerts negative radiation pressure.

\begin{figure}
 \begin{center}
 \includegraphics[width=1\linewidth]{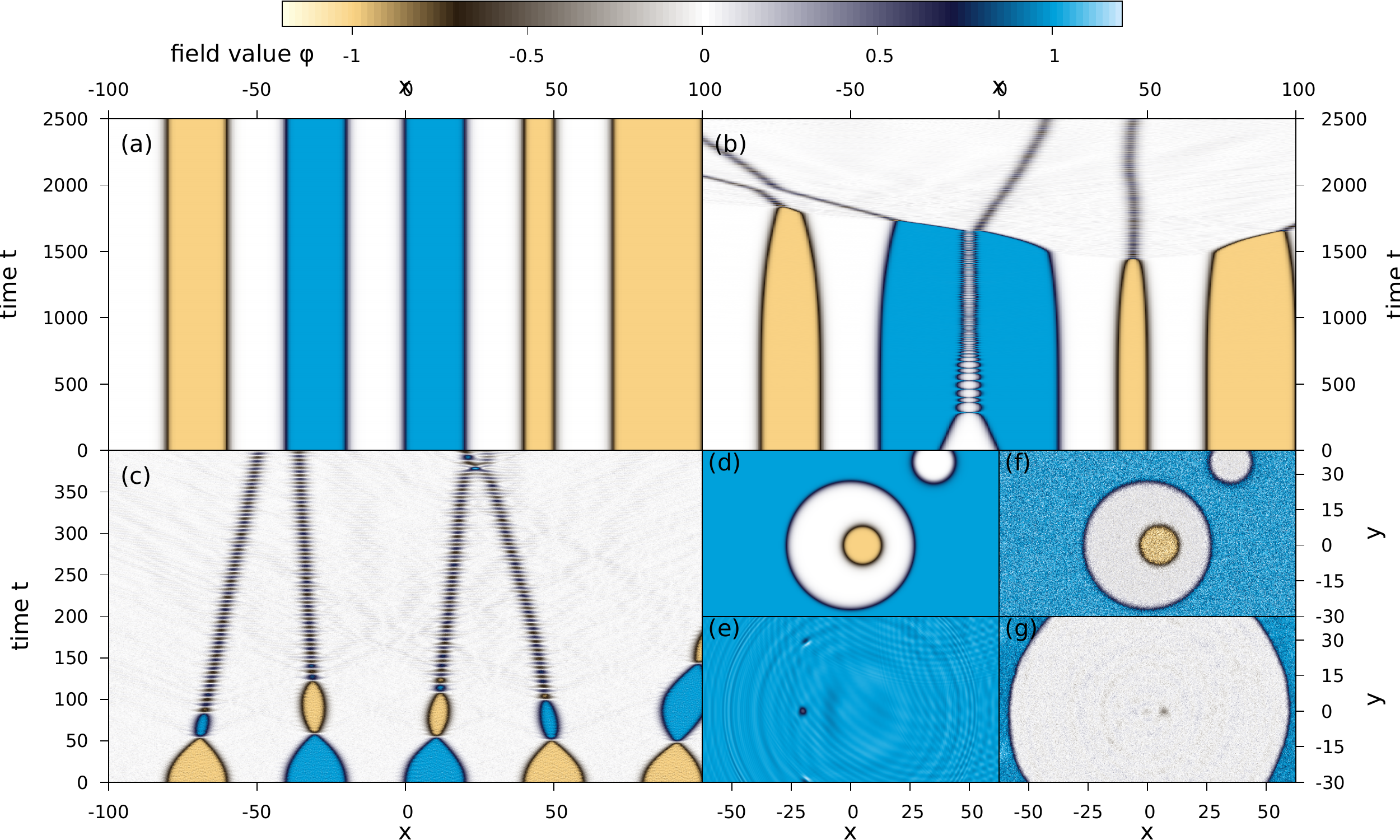}
 \caption{\small Evolution of system of kinks: (a) nearly static configuration, 
 (b) annihilation of close kink-antikink pair triggers the chain reaction of other annihilation, 
 (c) with gaussian 
noise with amplitude $A=0.05$. Evolution of two dimensional circular domain walls: without perturbation (d) $t=0$ (e) $t=75$ and with gaussian 
noise $A=0.12$ (f) $t=0$ (g) $t=75$.}\label{fig:2}
\end{center}
\end{figure}

{\it ~General perturbations.~}
Kinks interact very weakly with each other on large distances. Their profile vanish exponentially, and so do the interaction between them.
For a pair of kinks initially separated by the distance $L$ the estimated time to the collision is of order of $T\approx 2e^{L/2}/\sqrt{L}$. The 
value was numerically verified.
System of static, separated kinks can last even longer, because the forces from neighboring kinks cancel each other (Fig.~\ref{fig:2}.a, here 
for the first three kinks $L=20$ 
so the timescale to collision is about $10^4$). 
However, adding a small perturbation causes the radiation which exerts pressure on those kinks. We have tested this idea by adding a localized 
Gaussian profile $\phi=\phi_{kinks}+ae^{-bx^2}$ or colliding two kinks (Fig.~\ref{fig:2}.b). The collapse of the system was evidently faster compared 
to the case when no perturbation was added.
Because of the polarity in the direction of the radiation pressure,  the kinks always accelerate in such a way that the regions with vacuum $\phi=0$ 
grow and vacua $\phi=\pm1$ shrink. Moreover, when kink and antykink collide they form an oscillon. During such collisions more radiation is produced 
increasing the rate with which other kinks collide. Therefore a small perturbation can start a chain reaction leading to a collapse of the system of 
kinks.
We have also simulated this radiation by adding random noise $\langle\mathcal{A}\rangle\zeta(x)$ to the initial conditions, where 
$\langle\zeta(x)\zeta(x')\rangle=\delta(x-x')$ is a random number with Gauss distribution.
Adding such term is equivalent to introducing to the system background radiation which could be a relic of some phase transition 
(Fig.~\ref{fig:2}.c). Again, the system of kinks was quickly destabilized. 
We also solved Langevin equation but because of the damping term the dynamics was slowed down.
After collisions of kink and antikink an oscillon was formed. 
The oscillons are not stable but they live for a very long time. 
However, perturbation usually increase the rate with which the oscillon radiates shortening its life time. 
Therefore, initially almost static system of kinks can end up as almost uniform state filled with radiation around one stable vacuum. The 
time when this state is reached is determined by the life span of the oscillons in given conditions.
It is also worth mentioning that in some models collision of a kink and an antykink can lead to resonant windows and fractal structure. This happens 
usually when kinks have internal degrees of freedom (as in $\phi^4$) or if some meson states can be trapped between the kinks as it could happen in 
our model (bag model). However, the presented mechanism leads to collisions of such configurations which do not trap mesons.

{\it ~Other models.~}
Similar simulations for other models like $\phi^4$ revealed that the system of kinks can be destabilized by radiation, however, when the mass of 
small perturbation is the same around each vacuum there is no polarity in the direction of the acceleration of the kinks. The motion of kinks 
resembles a random walk.
In $\phi^4$  model the radiation from the collisions did not change the motion of the other kinks significantly. Kinks are transparent in the first 
order in the amplitude of the radiation. They undergo a negative radiation pressure which is proportional to the fourth power of the amplitude. Due 
the the nonlinear nature of this phenomenon, there is no simple superposition rule to sum all the effects coming from different frequencies. However 
we have found that the most dominant contribution to the motion of the kinks have collisions with oscillons created in earlier collisions. An 
oscillon can bounce back from the kink or go through it. However it can possesses large amount of energy which can be transfered to the kink, 
significantly increasing its velocity.

In the sine-Gordon model the kinks are completely transparent to the radiation due to the integrability of the model. During collisions no radiation 
is lost. Therefore the evolution of system of kinks is completely determined by initial conditions. No annihilation or creation is possible within 
the sG model.

In general, if all the vacua have not only the same value but also the same mass of small perturbations there is no difference as to the direction 
from which the radiation came. Contributions coming from both sides should be equal and cancel each other.
It does not matter whether the single, monochromatic wave traveling in one direction exerts a positive or negative radiation pressure. No vacuum is 
distinguished.
On the other hand if the masses are different, as they are in the discussed $\phi^6$ model, some excitations are more easily excited and can give 
nonvanishing contribution pushing the kink towards the vacuum around which the perturbations have larger mass.

It is also possible to construct more complicated theories including gauge fileds or supersymmetry. All the above consideretaions should hold only if 
there is a difference in masses of small fluctuations around different vacua.

{\it ~Effective model.~}
When the radiation fills uniformly the entire space we can integrate out small degrees of freedom and try to construct a more effective model 
describing only the motion of the kinks. 
Suppose that we know the spectral distribution of the radiation $\mathcal{A}(\omega)$ (which could be for instance the spectrum of the black body 
radiation). In $\phi^6$ the radiation pressure comes from the first order perturbation series, therefore we can use the ordinary superposition rule. 
The total force exerted by the radiation could be written as an integral over all frequencies.
\begin{equation}
   F_{tot}=\int_{-\infty}^\infty\!d\omega\; \mathcal{A}(\omega)^2f(\omega)=\alpha \langle \mathcal{A}\rangle^2.
\end{equation} 
where $\alpha$ is a value which in principle should depend on the distribution.
In case of uniform distribution for all frequencies we have two contributions corresponding to two waves coming from opposite directions. They should 
have the same amplitude, so they should add. The sum of those functions is almost constant (within 1\% for $\omega>2.2$) and equal to 1.5 so we can 
assume that $\alpha=0.75$. The discrepancy can be large only if low frequencies dominate. $\langle \mathcal{A}\rangle$ is the average amplitude of 
the perturbation. 
Note that all contributions $f(\omega)$ push the kink in the same direction (see eq. (\ref{eq:forza})).
Kinks separated by the distance $L=20$ as they are in FIG.~\ref{fig:2}c.~should collide after time $T=\sqrt{ML/F_{tot}}\approx 52$ which is in very 
good agreement with simulations.

Exactly the same force with just an opposite sign is exerted on the antinkink. This can be effectively described as if one of the vacua is 
raised.
It is quite easy to show that adding a term to the field theory potential, shifting vacua, results in external force acting on a kink, which in the 
first order can be written as 
$ F=-\Delta V/M$,
where $\Delta V$ is a gap between the vacua \cite{Shnir}.
Therefore the total force exerted by the radiation can be effectively written as a shift in the potential of the vacua $\pm 1$ by $\Delta V\sim\langle 
\mathcal{A}\rangle^2$.

\begin{figure}
 \begin{center}
 \includegraphics[width=1\linewidth]{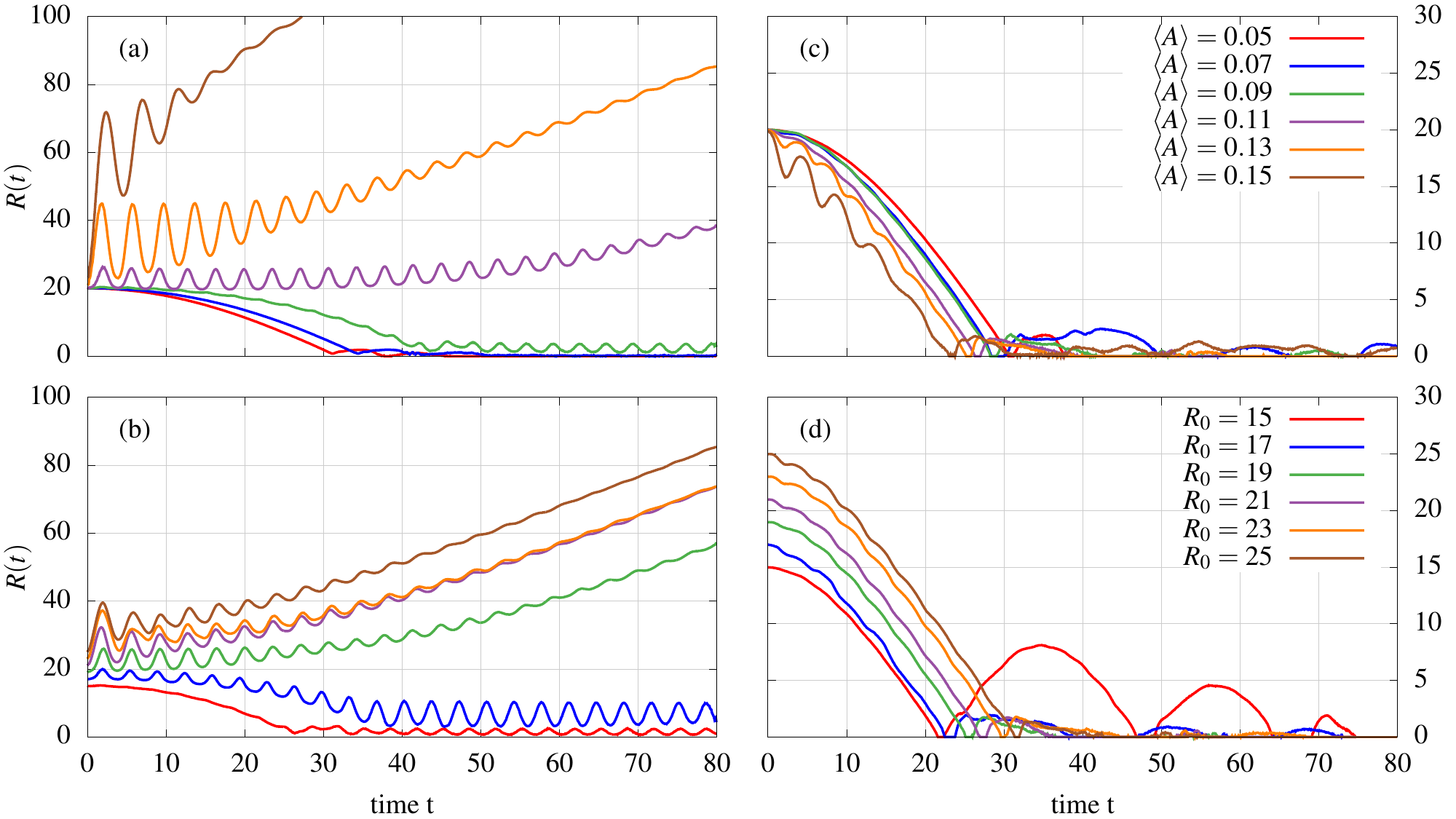}
\caption{\small A size of a bulk of vacuum in the presence of fluctuations. 
The upper plots (a) and (c) show the radius of the bulk for initial conditions $R_0=20$ and variable amplitude of fluctuations. The lower plots (b) 
and (d) show the evolution of the radius of bulks with different initial radii with the same fluctuation amplitude $\langle A\rangle=0.1$. 
Plots (a) and (b) correspond to enclosed vacuum $\phi_{in}=0$  and plots (c) and (d) to $\phi_{in}=1$.\\[-30pt]}\label{fig:3}
\end{center}
\end{figure}

{\it ~Two dimensional case.~}
Let us now consider domain walls in 2+1 dimensions. Effects described in the previous section are still present so the radiation pressure would try 
to close any bulks with vacua $\pm 1$ and bulks enclosing vacuum $\phi=0$ should grow. However, in two or more dimensions another phenomenon is 
present. 
The energy of a circular domain wall with large enough radius is equal to 
$ E(R)=2\pi MR $. This energy can be considered as the potential energy. The force acting on the unit length of the defect is equal to $F = -M/R$. 
More precise calculation of the evolution can be found for example in \cite{Arodz}.
If the field inside the domain wall reaches the value $\phi_{in}=\pm1$ and outside the wall $\phi_{out}=0$ the forces of radiation pressure and 
tension act in the same direction. So any 
radiation would speed up the contraction of such circles. On the other hand when $\phi_{in}=0$ and $\phi_{out}=\pm1$ the forces act in opposite 
directions. The tension can be balanced with the radiation pressure $F=\alpha\langle\mathcal{A}\rangle^2$, where $\alpha$ contains all the 
information about the distribution of the perturbations and the geometry of the defect. $\langle\mathcal{A}\rangle$ is a amplitude of the noise. 
Therefore there is a critical radius of the bulk $R_{crit}=\frac{M}{\alpha\langle\mathcal{A}\rangle^2}$. We have performed numerical simulations with 
initial conditions of a  circular domain wall with radius $R_0$ with additional random fluctuations of amplitude $\langle\mathcal{A}\rangle$. For 
given amplitude domain walls below certain radius shrank while domain walls which initially had a larger radius grew up (Fig.~\ref{fig:3} (a),(b)).
The simulations also confirmed our rough estimation that the critical radius indeed is proportional to $\langle\mathcal{A}\rangle^2$. From the 
critical values measured from simulations we can estimate that in the case of circular domain walls $\alpha\approx1.3$ which is almost twice the 
value obtained in one-dimensional case.

Domain walls enclosing $\phi=\pm1$ vacua shrank faster with larger amplitude of fluctuation (Fig.~\ref{fig:3} (c),(d)).
Moreover we have noticed that some radiation entering the bulk does not leave. After a while there is more radiative energy within the bulk. This 
creates additional pressure. Therefore initially static bulk starts to expand. However this effect is weak and does not influence much the above 
analisis.
Another observation we have noticed is that a number of oscillons was created. The oscillons were created in more or less the same phase. This fact 
was earlier noticed by Gleiser \cite{Gleiser}.

For comparison we have performed analogous numerical simulations in case of $\phi^4$ model. The radiation did not effect much the motion of the 
domain walls. Especially because the radiation coming from opposite sides of the wall exerts opposite effects which cancel each other out. In this 
case all the domain walls shrank with more or less the same rate.

In three dimensions the only difference is that the force shrinking the sphere is equal to $F=-2M/R$ which results in the critical radius being twice 
the critical radius in two dimensions. $R^{(3D)}_{crit}=\frac{2M}{\alpha\langle\mathcal{A}\rangle^2}$.

{\it ~Conclusions.~}
We have shown that in the $\phi^6$ model the differences in masses of small perturbations around different vacua cause the polarization of radiation 
pressure. No matter what is the characteristic of perturbation, the kinks would always accelerate in such a way that the vacua $\pm 1$ would 
disappear. 
The only stable vacuum is therefore $\phi=0$. This behavior is very similar to the effect when the vacua $\pm 1$ are raised. 
Therefore for uniformly distributed radiation one can introduce the effective model. 
The shift of the vacua can be expressed as an integral over all frequencies and is proportional to the square of the amplitude of the perturbation. 
Any radiation can force two kinks to accelerate towards each other. After the collision more radiation is created forcing other kinks to collide
faster. 
Therefore even small, local perturbation can start a chain reaction leading to the collapse of a system of initially almost static kinks. 
In higher dimensions, closed domain walls in the absence of other perturbation, usually shrink. 
The tension is proportional to the inverse of a radius. 

Radiation, or any type of fluctuations can speed up the process of domain wall decay, if vacuum with higher mass of small perturbation is closed 
inside.
If the radius of the domain enclosing the vacuum with smaller mass, the radiation can stop the decay and even accelerate the growth of such domain. 
The critical radius is proportional to the inverse second power of the average of perturbation amplitude.

The radiation pressure can be a cause of vanishing of kinks and domain walls in some models.  
This process is very rapid compared with the interaction of static topological defects. 
No such phenomenon was found in case where the vacua have the same masses. 

We are sure that the similar phenomenon should exist in more complicated models widely discussed in the literature \cite{Vachaspati}. The only 
requirement is that the field theory potential is not entirely symmetric although the energies of the vacua are the same. Scalar field coupled to a
gauge field can play a role of a Higgs field. Different expectation values of the Higgs field in different vacua can result in different masses of 
the gauge fields. The mechanism described in the present paper would favor the vacua with the smaller masses.  It is also possible to consider 
higher topological defects such as vortices or monopoles. Such defects in a presence of radiation would accelerate more likely in certain directions 
pointing towards the smallest mass of the field. The collisions with other defects on these directions would be more likely. 

{\it ~Supplementary material.~} The letter is complemented with simulations showing some of the features discussed.

\begin{small}

\end{small}
\end{document}


\section{Supplementary material - description of the attached movies}
\begin{enumerate}
 \item {\verb|1dWave.mov|}: Time dependent solution of the full PDE of the $\phi^6$ model with three different initial conditions each with two kinks 
and a wave coming from rhs. The radiation exerts negative radiation pressure on 
kinks supported on $\phi=\pm 1$ vacua and antikinks supported on $\phi=0$ vacuum. As a result the system evolves in such a way that the vacuum at 
$\phi=0$ grows. In the first case an oscillon remains as a relic of collision. In two other cases the kinks just go away in opposite directions.
\item \verb|2d_A000.mov|: Collapse of the circular domain walls in 2d in the absence of perturbation. As a result some radiation and oscillons remain.
\item \verb|2d_A012.mov|: The similar setup to the previous case but with additional random fluctuations added as initial conditions 
$\langle\mathcal{A}\rangle=0.12$. The circle containing vacuum $-1$ collapse even faster than in the absence of radiation. Small circle with the 
vacuum $0$ also collapse but slower as the fluctuations cannot balance the tension. The large circle with supercritical radius grows expanding the 
vacuum $\phi=0$. The size and resolution of movie was reduced for easier downloads.
\end{enumerate}